\documentclass[journal=jpclcd,manuscript=article, email=false]{achemso}

\usepackage[version=3]{mhchem} 
\usepackage[T1]{fontenc}       
\usepackage{graphicx}

\usepackage{color}

\usepackage{soul}
\usepackage{gensymb}

\author{R. Pasquier}
\affiliation{Institut de Physique et de Chimie des Mat\'eriaux de Strasbourg (IPCMS), Universit\'e de Strasbourg, CNRS UMR 7504, 23 rue du Loess, BP 43, F-67034 Strasbourg Cedex 2, France}

\author{K. Rassoul}
\affiliation{Institut de Physique et de Chimie des Mat\'eriaux de Strasbourg (IPCMS), Universit\'e de Strasbourg, CNRS UMR 7504, 23 rue du Loess, BP 43, F-67034 Strasbourg Cedex 2, France}
\author{M. Alouani}
\affiliation{Institut de Physique et de Chimie des Mat\'eriaux de Strasbourg (IPCMS), Universit\'e de Strasbourg, CNRS UMR 7504, 23 rue du Loess, BP 43, F-67034 Strasbourg Cedex 2, France}
\email{mea@ipcms.unistra.fr}

\title{ Inverse spin crossover in  fluorinated Fe(1,10-phenanthroline)$_2$(NCS)$_2$ adsorbed on Cu (001) surface}
\begin{document}

\setlength{\fboxrule}{0 pt}
\begin{tocentry}

\begin{center}
\includegraphics{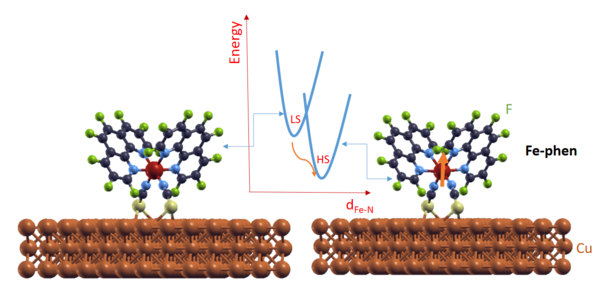}%
\end{center}

\end{tocentry}


\newpage

\begin{abstract} 
Density functional theory (DFT) including van der Waals weak interaction in
    conjunction with the so called rotational invariant DFT+U, where $U$ is the
    Hubbard interaction of the iron site, is used to show that the fluorinated
    spin crossover Fe(phen$)_{2}$(NCS$)_{2}$ molecule whether in the gas phase
    or adsorbed on Cu(001) surface switches from the original low spin state to
    the high spin state.  Using Bader electron density analysis, 
    this inversion of the spin-crossover is explained in
    terms of electron doping of the Fe-octahedron cage which led  to an increase of the Fe-N bond lengths.
    Consequently, the ligand-field splitting is drastically reduced, making
    the high-spin ground state more stable than the low-spin state.  The calculated
    scanning tunneling microscopy (STM) images in the Tersoff-Hamann
    approximation show a clear distinction between the fluorinated and the
    unfluorinated molecule.  This theoretical prediction is awaiting  future STM
    experimental confirmation. 
\end{abstract}

\maketitle

\section{Introduction}
The interest in spin crossover (SCO) molecules\cite{Cambi} is due to their
ability  to switch from the $^1A_1$  low spin (LS) state at low temperatures
to the $^5T_2$ high spin (HS) state at higher temperatures. The
transition between these two states is accompanied with a change in their
structural, electronic and magnetic properties.  The ease and
multitude ways leading to the spin crossover, such as the variation in temperature, light,
pressure, magnetic field, or electric field, make these molecules 
potential candidates for applications in spintronics and magnetic
recording \cite{Wolf,theseDavesne}. However the high sensitivity of the SCO
molecules to their environment makes their control challenging \cite{Davesne2012},
especially when they are in interaction with a substrate, where recent
experiments showed a fragmentation and distortion of many of these
molecules \cite{Knaak2019}. However, recently the principle of electrical 
detection of the light-induced spin transition in SCO/graphene heterostructures was 
demonstrated \cite{Jeff1} and even room temperature optoelectronic device operating with spin
crossover nanoparticles\cite{Jeff}.

The SCO molecules are
composed essentially of a transition element ion complex, like iron (II), surrounded by
ligands. When the ligands take an octahedral or tetrahedral conformation around
the metallic element, they generate a field called "ligand field $\Delta$"
operating on the $d$ orbitals of the metallic ion. This causes a partial lifting of
degeneracy in the $d$-orbitals leading to the splitting of the $ e_{g}$ and $t_{2g}$
energy levels by $\Delta$. The electronic configuration of these two
levels and hence the SCO ground state depends on the competition between the ligand
field $\Delta$ and the Coulomb electronic repulsion $U$ between the $d$ electrons. 
If $\Delta$ is smaller
than $U$ the high-spin state is favored, on the other hand if $\Delta$ is larger
than $U$ the low spin-state is favored.

 One of the most interesting challenges is to confirm whether these spin-crossover molecules can
 maintain their LS-HS  spin-crossover when deposited on  a metallic substrate.
 Previous experimental works \cite{theseDavesne,Davesne2012,Gopakumar2012} 
 on individual Fe(phen$)_{2}$(NCS$)_{2}$ molecules \cite{Baker1964}
 (labeled Fephen) on   Cu(001) or Au(111) identified both magnetic
 conformations when the surface is  scanned by scanning tunneling microscopy
 (STM), showing two different sizes of the double  phen lobes corresponding respectively
 to the HS and LS states. Distinction between 
 the spin state of each conformation was done by scanning tunneling spectroscopy (STS),
 observing a Kondo peak in the high-spin (HS) state  whereas only a low
 electric conductivity was obtained in the LS state. 
 The screening of the iron spin magnetic moment was attributed to the interface conduction electrons\cite{theseDavesne,Davesne2012}. 
 However, unsuccessful efforts  have been made to switch the molecules  on the surface by applying an electric
 field, high tunneling voltage or high current density \cite{theseDavesne,Davesne2012}. 
 It was concluded that the molecules are trapped in their magnetic states, LS or HS, 
 due to the strong interaction between the two NCS groups of the molecule   and the
 metallic substrate via the sulfur atoms. However, the introduction of a monolayer of CuN between the molecule and 
 the copper surface led to a
 reversible electrical switching of the Fephen molecule from the  LS to the  HS state. 
 This is corroborated by theoretical results, which showed a significant reduction of the difference of 
 the adsorption energy between the two magnetic states  of the
 Fephen molecule adsorbed on a CuN(100) substrate \cite{Gueddida2013}. 

To control the HS-LS transition of  
the SCO molecules on metallic surfaces,  it is desirable to find a way to control or tune 
the ligand field and hence the spin
crossover energy barrier. In this work our aim is therefore to perform such a tuning of the ligand field by
fluorination of the molecule, and  thus creating a perturbation on the octahedral ligands. 
We can therefore monitor the evolution of the ligand field 
and the energy barrier between the
two magnetic states of the molecule as a function of this perturbation. 
To this end, we computed the minimum energy path
(MEP) between the two HS and LS ground states when the iron ligands of the spin-crossover
molecule were modified by fluorination. We show that fluorination 
can invert the molecule spin state, i.e., making the HS state lower in energy than the natural low temperature LS
state of the molecule. Thus our results demonstrate that he HS and LS adsorption energy
can be tuned by choosing an appropriate dopant to the molecule.  To make the calculation, we have chosen the  
Fe(Phen)$_{2}$(NCS)$_{2}$ molecule (Phen = 1,10-phenanthroline) (see Fig. \ref{fig:relaxed}) 
and substituted  all the hydrogen atoms by fluorine atoms. Using Bader analysis,  we show that the difference of
electro-negativity between fluorine and carbon  leads to a reorganization of
electrons around the iron octahedron which modifies its  ensuing physical properties.  
For simplicity, we label H-Fephen  the unfluorinated molecule, and 
F-Fephen the fluorinated molecule.

 The paper is organized as follows. In the second section we provide our method of calculation and give the most
 relevant parameters needed to perform the electronic structure and compute the total energy. 
 In the third section we describe the effect of fluorination
 on the spin crossover molecule,  analyze the electron distribution using Bader analyses  and 
 show its effect on inversion of the spin crossover. We then analyze the iron density of states in terms of a
 simple ligand-field charge model for both the unfluorinated and the 
 fluorinated molecule. In the fourth section we compute the
 minimal energy path by constrained minimization and the so called nudged elastic band method (NEB) and in the fifth
 section we report and analyze the constant current STM images. 

\section{Effect of fluorination on the spin-crossover molecule}
\subsection{Adsorption of F-Fephen on Cu (001)}
The H-Fephen molecule is one of the most interesting iron (II) spin crossover molecules,
discovered by Baker in 1964 \cite{Baker1964}. In this molecule, the iron(II) atom is
surrounded by 6 nitrogen atoms forming an octahedral shape. The HS-LS transition 
in this molecule leads essentially to the  variation of the bond lengths and bond angles
between the iron and nitrogen atoms and we observe a significant increase  of these bond lengths
in the HS state compared to the LS state.



By substituting hydrogen by fluorine,  we obtain the fluorinated molecule
(labeled F-Fephen).  After atomic relaxations,  except for some elongation in the
iron-nitrogen bond lengths, the global geometry of the  structure of the molecule
for the two magnetic states remain similar to  those of the unfluorinated molecule. However, we
will show that these elongations are of  crucial importance, as they lead to an
inversion in the total energy  of the two magnetic states of the molecule, i.e., the
high-spin state (HS) is now the most stable state.
We report in table \ref{tab:table} the energies and magnetic moments obtained for 
the relaxed free unfluorinated and fluorinated molecule in the two LS and HS magnetic states and 
the difference of the total in kJ/mole which shows this inversion.
%
%

\begin{figure}[!ht]
	\centering{ \includegraphics[scale=0.440]{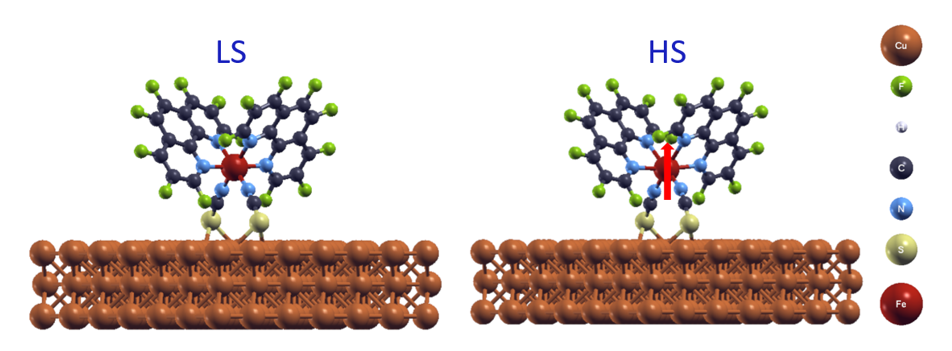}  }
    \caption{Relaxed LS (left) and HS (right) structures of the F-Fephen molecule on Cu(001)  substrate.}
    \label{fig:relaxed}
\end{figure}

The F-Fephen molecule adsorbed on copper surface
shows the same behavior as the unfluorinated molecule. Figure \ref{fig:relaxed} shows the relaxed 
structure of the two LS and
HS magnetic states. The equilibrium position is the same as the unfluorinated 
molecule, i.e ; the sulfur atom is in the bridge position and is 1.97 \r{A} away from
the  substrate in the LS state, compared to 1.95 \r{A} in the high-spin state in agreement with 
the results of Ref. \cite{Gueddida2013}.

\begin{figure}[!ht] 
	\centering{ \includegraphics[scale=0.15]{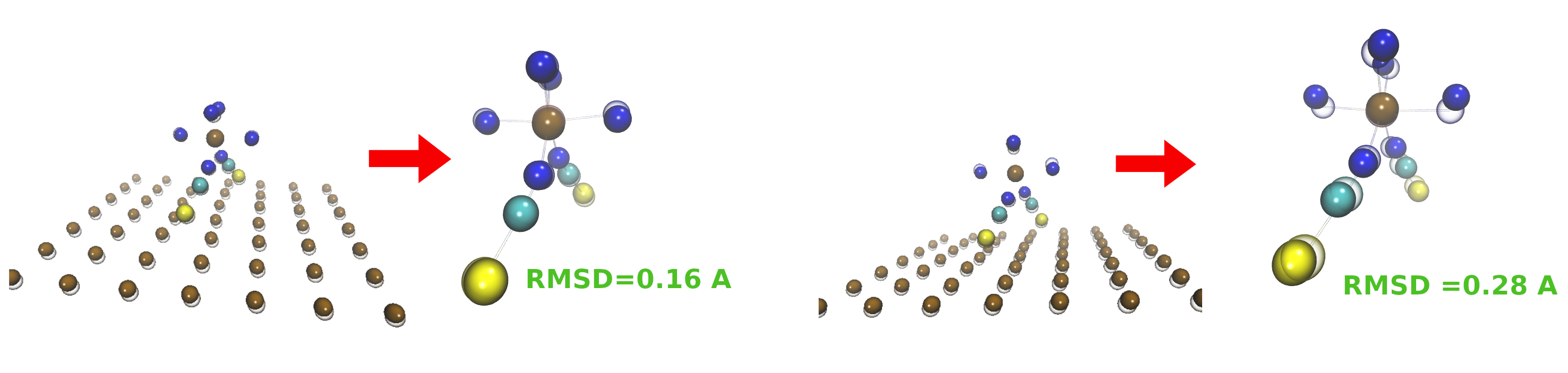}  }
\caption{Superposition on the iron atom of the octahedral and (NCS) groups 
of the relaxed H-Fephen and F-Fephen molecules on the substrate in the two
magnetic state (LS-left, HS-right), in color for the F-Fephen, and transparent for the H-Fephen.}
\label{fig:superposition}
\end{figure}

As in the free molecule case, the global molecular structure remains similar 
and the fluorination causes an elongation in the iron-nitrogen bonds length. 
We report in Figure \ref{fig:superposition} the
superposition of the H-Fephen and F-Fephen molecule at the octahedral and
NCS level, LS left and HS right. We observe  that the fluorination causes
significant structural alterations of the octahedral environment of the
molecule. The average root mean square deviations (RMSD) of the bond lengths show that this structural
deformation is more important in the HS state.  The two magnetic states are still conserved
when the molecule is adsorbed in the substrate, however, similar to the free molecule, 
 the total energy is lower for the HS state. We  report in table \ref{tab:table}
 the total energies and total magnetic moment for the molecule adsorbed on Cu(001) for the 
 two LS and HS magnetic states. However, it is important to mention that the fluorination 
 without relaxation of the molecule did not cause the spin crossover inversion. Indeed, although
 the HS-LS difference of total energy is significantly reduced from 55.3 kJ/mole in the unfluorinated case
 to 20.4 kJ/mole, it is only when
 the atoms of the molecule  are allowed to relax that this difference becomes negative 
 as shown in table \ref{tab:table}. This underlines
 the importance of atomic relaxation influenced by the relatively strong  electronegativity of the fluorine atom 
 compared to that of hydrogen.
 The effect of fluorination on the atomic relaxation will be further studied in the next subsection by analyzing
 the electronic distribution on each atom by means of Bader  analysis\cite{Henkelman2006}.
\begin{table}[H]
	\centering{
\begin{tabular}{l l r r r r r}
\hline
\hline
 &    &&  LS              &    HS                    && $\Delta E$ (kJ/mole)   \\
\hline

	Unfluorinated     &$E_T$ (eV)     && -749.57 (-359.726)  &   -749.00 (-359.517)    &&  55.3 (20.3)     \\
	& $\mu_{\rm Fe}$ ($ \mu_{\rm B} $) &&  0.00 (0.00)       &   3.60 (3.72)             & &    \\  
	Fluorinated     &$E_T$ (eV)     &&  -887.407 (361.701)   &   -887.609 (361.977)    &&  -19.6 (-26.8)     \\
	& $\mu_{\rm Fe}$ ($ \mu_{\rm B} $) &&  0.00 (0.00)       &   3.25 (3.72)             & &    \\  
\hline
\hline
\end{tabular}
}
\caption{ Total energy ($E_T$) in eV and iron magnetic moment ($\mu_{\rm Fe}$) in Bohr magneton ($ \mu_{\rm B} $) 
for the high spin (HS) and low spin (LS) states of  the unfluorinated and fluorinated Fephen molecule 
	adsorbed  on the Cu(001) substrate (the values for the free molecule are between parenthesis). 
	The last column shows the total energy difference $\Delta E = E_T^{\rm HS} -E_T^{\rm LS}$ in kJ/mole.}
\label{tab:table}
\end{table}  
\subsection{Electron distribution analysis.} 
\begin{figure} [!ht]
      \begin{center}
\includegraphics[scale=0.2]{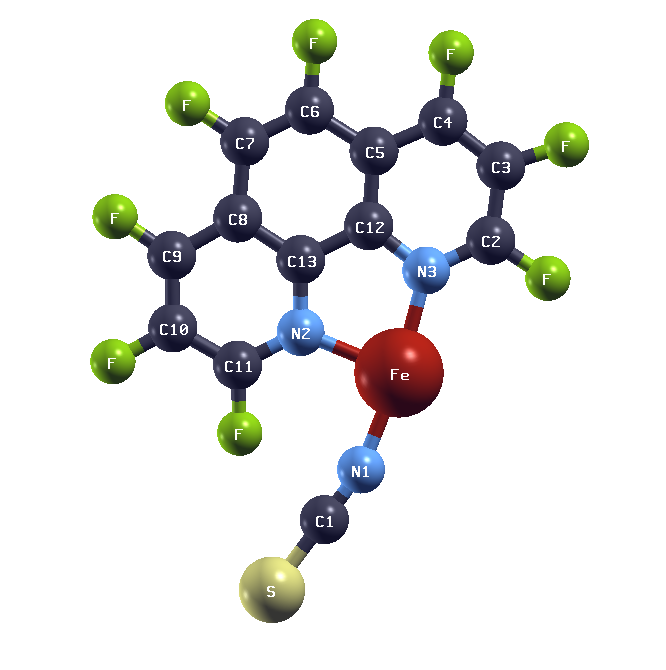}
      \end{center}
    \begin{minipage}{0.5\linewidth}
        \includegraphics[scale=0.37]{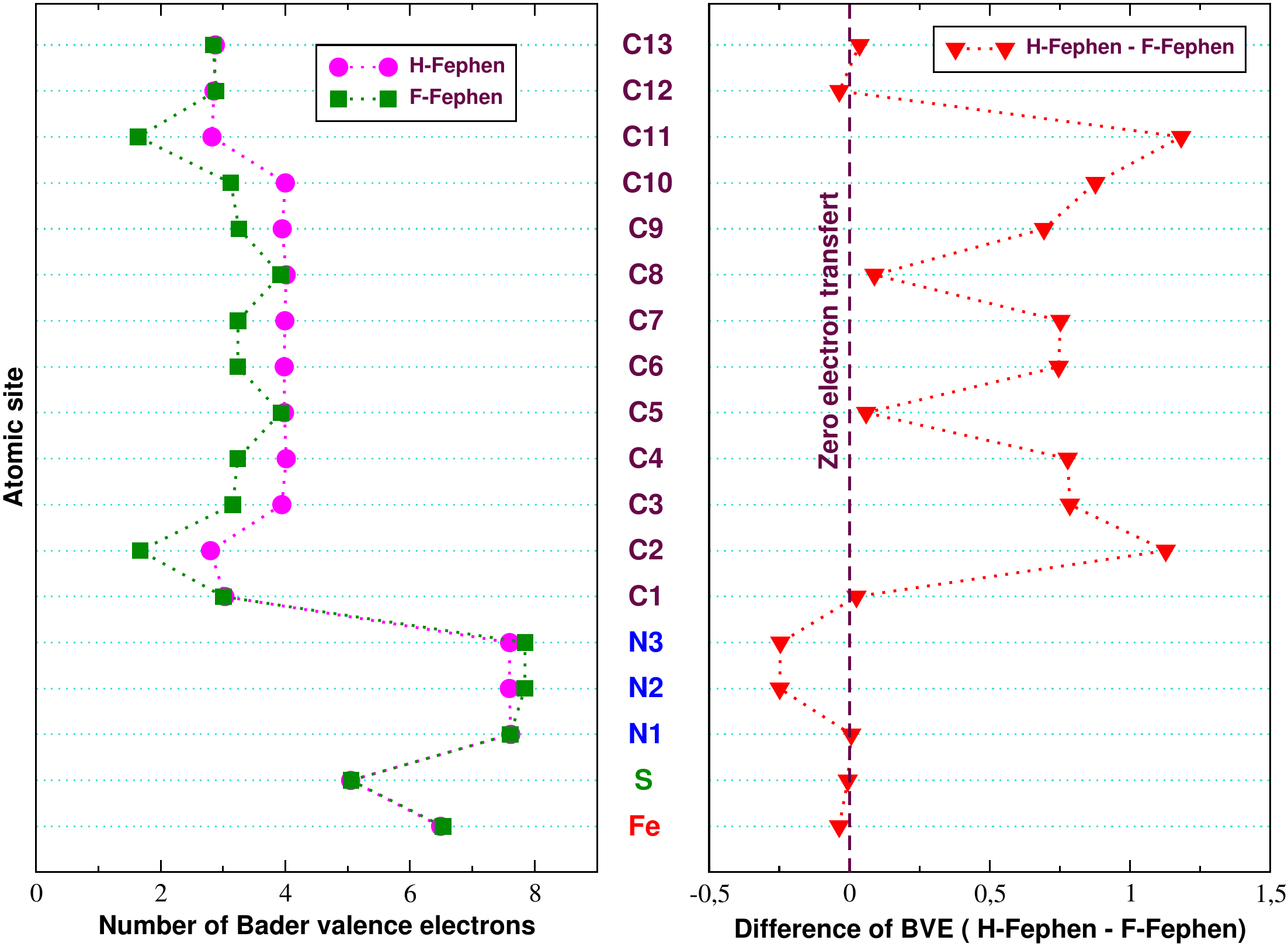} %
    \end{minipage}%
    \begin{minipage}{0.5\linewidth}
        \includegraphics[scale=0.37]{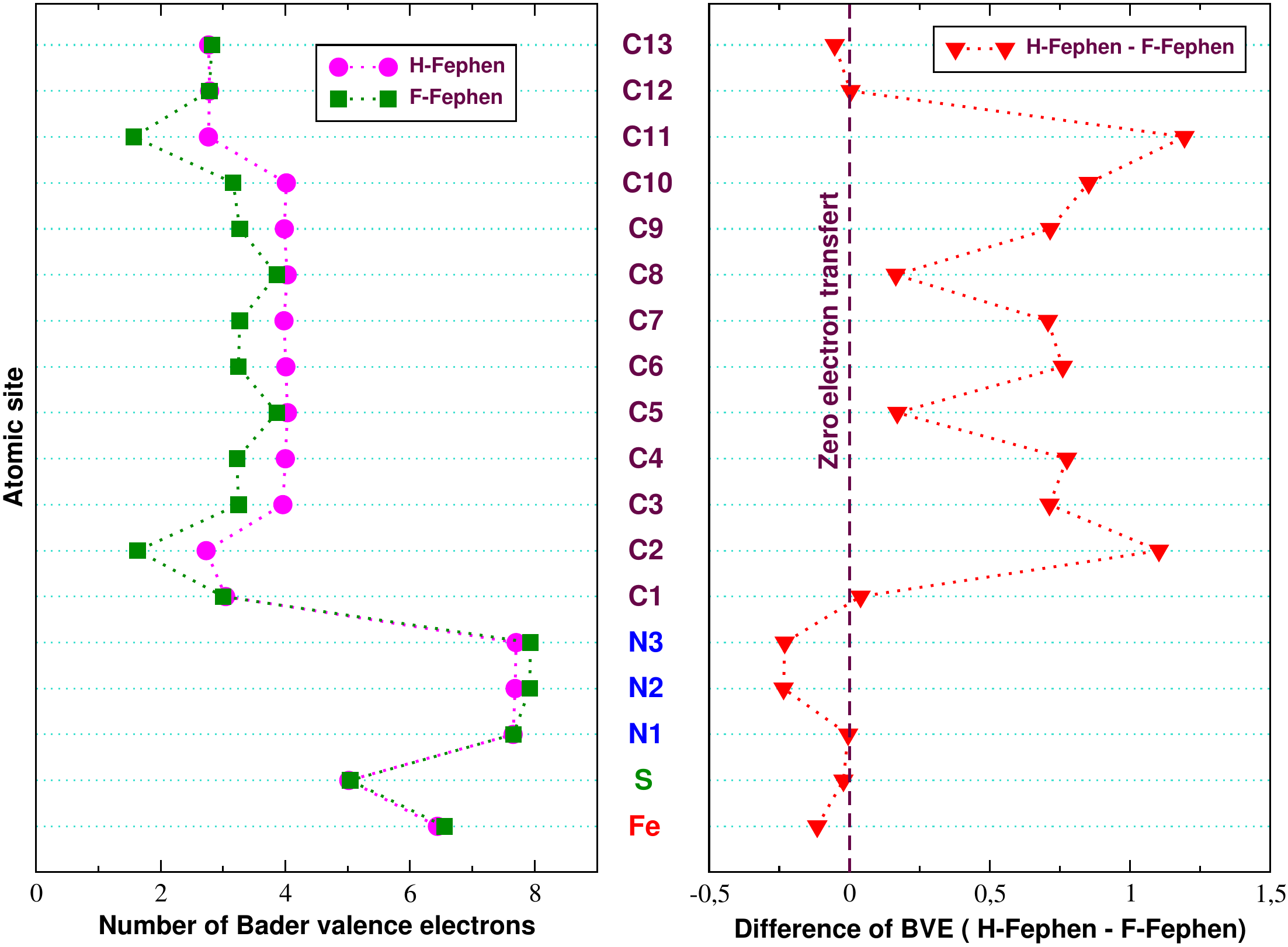} %
    \end{minipage}
	  \caption{ From left to right: (1) Valence Bader electrons (VBE) per atom 
	   for  H-Fephen (pink dots), F-Fephen (green squares),
	  (2) LS VBE difference between H-Fephen and F-Fephen (red triangles) for each site, (3) and (4) are 
	  respectively same as for (1) and (2) but for HS state. 
       The labels of atoms are as shown  on the top picture of the moiety  of the molecule.}
      \label{fig:bader}
      \end{figure}

It is important to determine the effect of the distortion of the iron
octahedron on the electronic levels of the molecule due to fluorination. 
Using Bader analysis \cite{Henkelman2006},   we focus on the electron distribution for each atom and its
variation as we pass from the H-Fephen to F-Fephen. Fig. \ref{fig:bader} shows the Bader valence
electron difference for each atom of the H-Fephen and F-Fephen molecule
adsorbed on Cu(001) surface. Due to the relative symmetry of the
Fe(Phen)$_{2}$(NCS)$_{2}$ molecule, we consider only one phen group.
We then obtain for each atom the number of valence electrons before (pink dots) and after fluorination (green
squares). 
To show the electron transfer caused by fluorination, we calculated the electron difference in
each site between the H and F-Fephen. This is shown by a red triangles in Fig. \ref{fig:bader}   for
each magnetic state.

The electron transfer occurs mainly in the carbon atoms 
which have a direct bond with fluorine. This transfer is in general uniform for
those atoms except for the C2 and C11, with $ 1.14$ to $1.28 $ electron in the
LS state and $ 1.24$ to $ 1.23$ electron in the HS state. Indeed we notice an
additional transfer from the two atoms. This excess of electrons is captured by
the two nitrogen N2 and N3, and this situation is observed in the two magnetic
states.
The excess electrons on the two nitrogen atoms of the phenyl group increases the
electrostatic repulsion with the iron site. This repulsion leads to the
elongation of the bond length between the central  iron and the neighboring
nitrogen atoms,  especially in the phen group. The bonding energy being lower
in the HS state compared to the LS state, the former is subject to more
distortion in the octahedral environment as it was already seen in the
previous subsection.

As a result of
the elongation of the Fe-N bonds, the octahedral ligand field is reduced and
the Coulomb repulsion among the  iron $d$ electrons becomes larger 
because of the reduced ligand screening. This leads to the stabilization of the
high-spin over the LS state in comparison to the unfluorinated case.  

\subsection{Iron  density of states and ligand-field analysis.} 
\begin{figure}[!ht]
	\centering{ \includegraphics[scale=0.5]{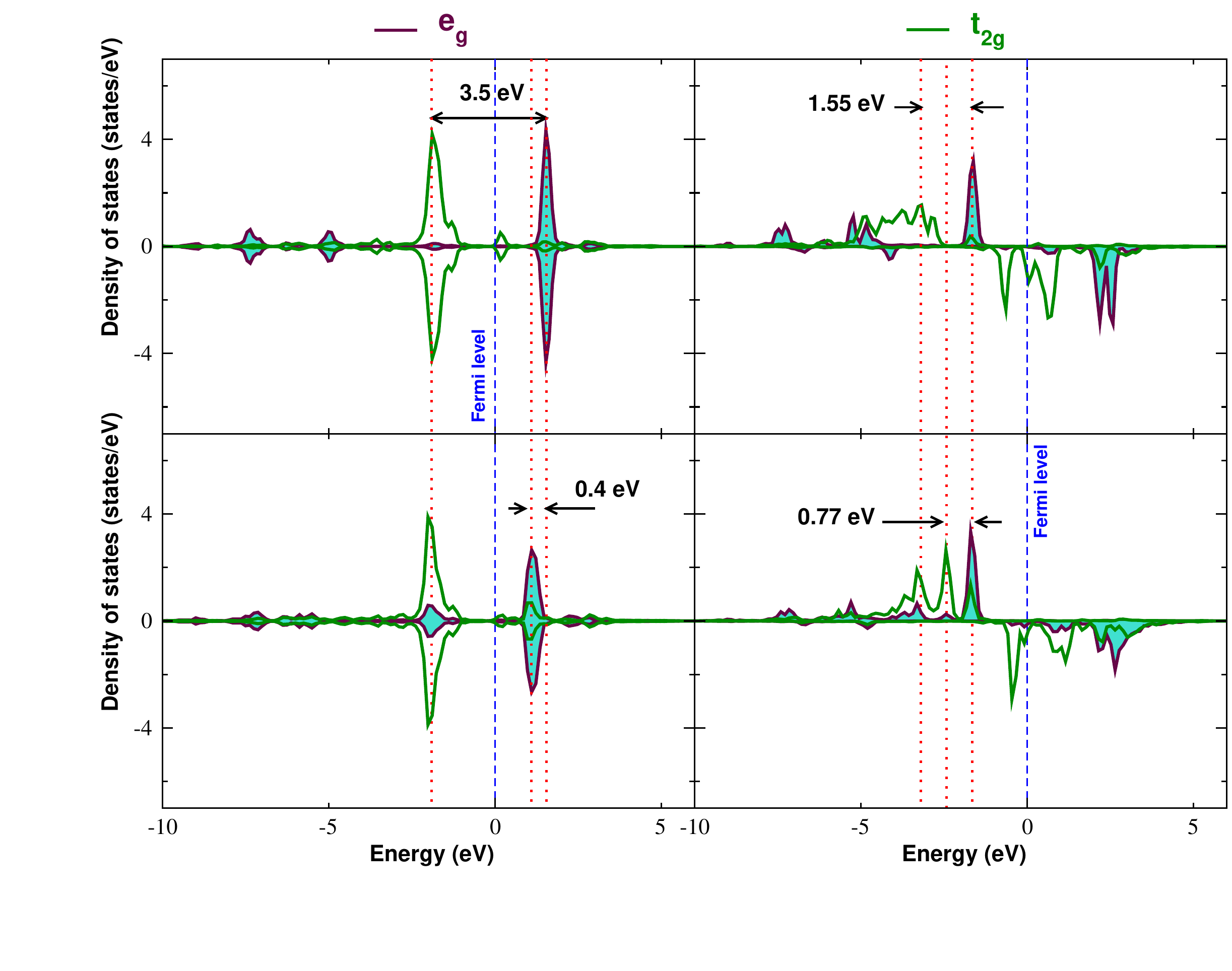} }
\caption{LS (left), HS (right) iron $d$-DOS for the Fe(Phen)$_{2}$(NCS)$_{2}$ molecule 
	on Cu(001) surface, H-Fephen (top) and F-Fephen(bottom). The $e_g$ states are shown
	in brown and filled green and $t_{2g}$ states  in green. The red vertical doted lines
	show the alignments of states and the reduction of the ligand field. The Fermi energy level is 
	at zero energy.}
\label{fig:dos}
\end{figure}

It is now interesting to see how the substitution of the hydrogen atoms by
fluorine atoms affects the iron density of states. Figure \ref{fig:dos} shows
how the $e_{g}$ and $t_{2g}$ ligand-field splitting is modified by fluorination.
For the unfluorinated molecule the ligand-field splitting in the low-spin state is
$3.50$ eV compared to $1.55$ eV in the high-spin state. For the F-Fephen, the
ligand field in both spin states is significantly reduced to $3.1$ eV in the
LS state and $0.77$ eV in the HS-state. In the LS state, this reduction is due 
to the shift of the $e_{g}$ level towards lower energies, and in the HS state
to the shift of the $t_{2g}$ towards high energies. As stated earlier, this
reduction of ligand-field splitting and the increase of the pairing Coulomb
energy among the $t_{2g}$ electrons lead to the stabilization of the high-spin
over the LS state.
The ligand-field theory helps describe the splitting of the electron energy states 
of Fe in the Fephen molecule, in the presence of a Coulomb potential generated
essentially by the first neighbor nitrogen atoms. This model has been very
successful in explaining the  spin-crossover phenomenon, where the ligand-field energy 
can be of the order of the pairing energy between electrons\cite{Ballhausen}.  

In the case of an octahedral environment, the electrons of the nitrogen are
closer to the the $d_{z^2}$ and $d_{x^2 - y^2}$ orbitals of Fe and away from the
$d_{xy}$, $d_{yz}$, and $d_{xz}$ orbitals. The strong electrostatic between the
atoms of nitrogen and iron  would lead to a splitting of the five $d$ orbitals
into two subshells, called $t_{2g}$, consisting of the $d_{xy}$, $d_{yz}$, and
$d_{xz}$ orbitals, and $e_g$ consisting of the $d_{z^2}$ and $d_{x^2 - y^2}$
orbitals, with the former being lower in energy due to a lower electron
repulsion compared to the latter. 
Two important details should be pointed out here: first, that the 
$e_g$ and $t_{2g}$, 
orbitals are described using real spherical harmonics, also known as
cubic harmonics, and second, that these are defined through their $m$ quantum
number, i.e., the eigenvalue of the $L_z$ operator, and depend on the
coordinate system.  To obtain the effect of the ligand field on the 
$e_g$ and $t_{2g}$ DOS, we need either
to rotate the real spherical harmonics from the global coordinate system to the
local one.  We have implemented this feature in \textsc{VASP}, the details are given 
in Ref. \cite{Dixit}.  We present below our results for the octahedral cationic Fe site in
Fephen.


The iron $e_g$ and $t_{2g}$ 
DOSs are shown respectively in brown  and green in Fig. \ref{fig:dos}. We observe
that the occupied 3$d$ states are clearly split,  with the 
 $t_{2g}$ orbitals lower in energy than the $e_{g}$, and that the ligand field is much
 reduced for the fluorinated molecule by 0.4 eV for the LS state and about 0.8 eV for the HS state.
 Using the ligand-field splitting, 
it is possible to probe the filling of $d$
orbitals and understand why the spin magnetic moment of Fe is about  
4$\mu_B$ as shown in the DOS given in  Fig.  \ref{fig:dos}.
To understand the splitting in an octahedral environment, a point-charge model
was developed, in which the effects of hybridization can be fully neglected.
Such a model would help us understand qualitatively the 
$e_g$ and $t_{2g}$ splitting. The
electrons in the model can only interact through the Coulomb interaction with
the neighboring negatively charged nitrogen ions. With this requirement, 
 we assumed a formal ionic charge of +2 on the central atom and -2
distributed on the nitrogen neighboring atoms forming the octahedron. To
maintain consistency between the ab initio calculation and the point-charge
model, the octahedron was oriented in the same direction as in the rotated frame of
reference, i.e., with the octahedral arms aligned maximally along the axes the reference frame. 
The Hamiltonian describing the ligand field is given by
\begin{equation}
	\hat H^{\sigma}_\lambda = V + \frac{I_\lambda}{2}  \sigma_z, 
\end{equation} 
where, $\lambda$ is set to
the LS or the HS state,  $I_\lambda$ is the an effective Stoner parameter
splitting of the spin up and spin down DOS for the HS state and is about 3 eV as
given by our ab initio calculation   and is  zero for the
low spin state, and $\sigma_z$ is the $z$  component Pauli matrix.
The potential $V$ due to the nitrogen ligands 
and felt by the central atom is given by 
\begin{equation} V(\boldsymbol{r}) =  \sum\limits_{i=1}^6\frac{q_i}
{|\boldsymbol{r}-\boldsymbol{R}_i|} = 4 \pi\sum\limits_{i=1}^6
q_i\sum\limits_{\ell=0}^\infty \sum\limits_{m=-\ell}^\ell
    (-1)^m\frac{1}{2\ell+1}\,\mathcal{Y}_\ell^{-m}(\hat{\boldsymbol{r}})\,
    \mathcal{Y}_\ell^m(\hat{\boldsymbol{R}}_i)\,\frac{r^\ell_<}{r^{\ell+1}_>},
\end{equation} 
where  $r_<$ ($r_>$) is the smaller (larger) radius between $r$ and $R_i$
and where $q_i = C e^2$, $C$ being  an effective product of  charges in each
nitrogen atom and the central iron atom, the $\mathcal{Y}_\ell^m$ are the
spherical harmonics, and $\boldsymbol{R}_i$ are the distance vectors connecting
the central iron atom to the different ligands $i$.  Note that the local reference frame
is centered on the iron atom and its axes are along the ligands if the
octahedron is perfect.  Since the octahedron is deformed, the local frame is
optimized to reduce the angles between $\boldsymbol{R}_{i}$ and the corresponding axis.
The spherical angles $\vartheta$ and $\varphi$ of the vector $\boldsymbol{R}_{i}$ are then
obtained in this optimized reference frame to compute the spherical  harmonics.
\begin{figure} [!ht]
\centering{
\begin{tabular}{c c}
	\includegraphics[scale=0.42] {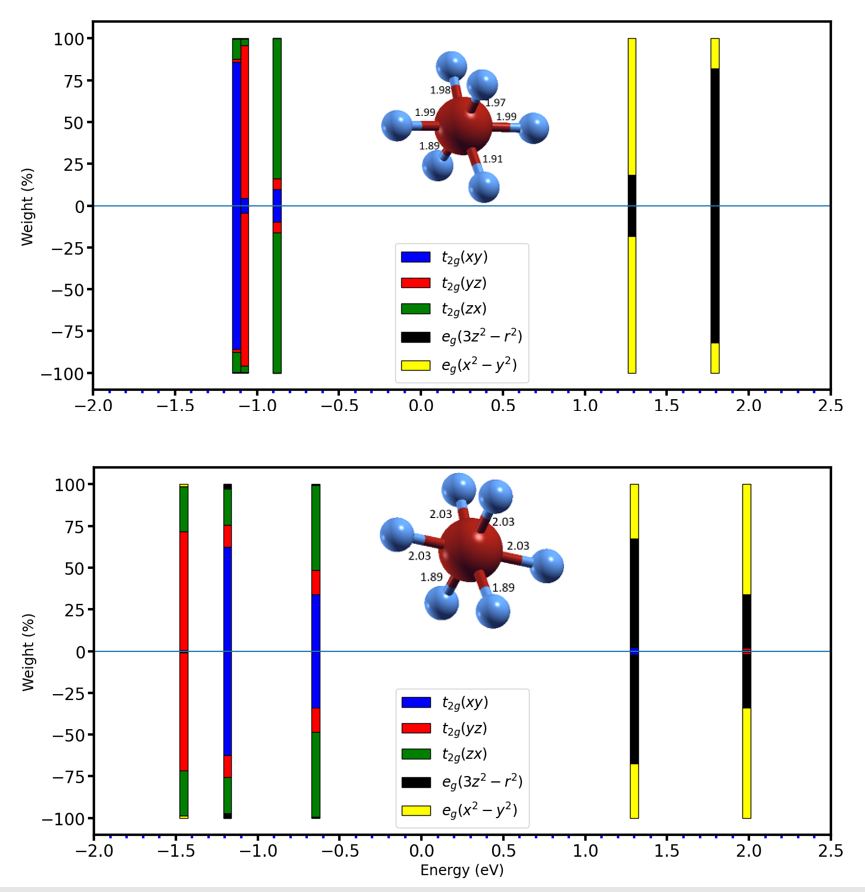}  &
         \includegraphics[scale=0.42]{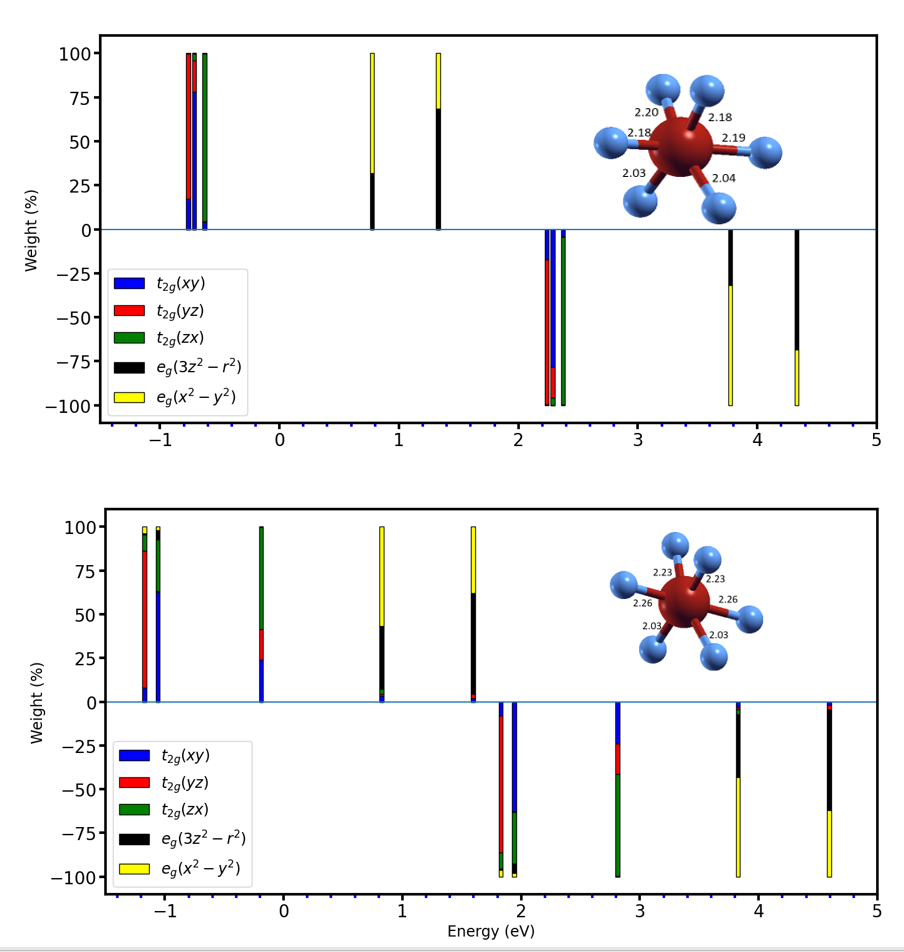} 
\end{tabular}
    }
\caption{
Ligand-field splittings using the point-charge model for both the LS (a) and HS (b) state and 
for the unfluorinated (top plots) and the fluorinated (bottom plots) Fephen. 
The distorted octahedron obtained from ab initio calculation are shown in the insets 
    (the bond lengths are in \AA).  The color code shows the different symmetries of the eigenvectors.
}
\label{fig:pointcharge}
\end{figure}
The results of this point-charge model are shown in Fig. \ref{fig:pointcharge} for
both LS and HS and for the unfluorinated  and fluorinated Fephen. The geometries of
the octahedron are taken from our ab initio relaxed structures on Cu(001) and are
shown in the insets of the figure. The model clearly shows that the unfluorinated molecule has 
a much higher ligand-field splitting compared to the fluorinated molecule in both the LS
and HS states. However, the splittings are much smaller than the ab initio results. The model also
shows that due to the distortion of the octahedron, the degeneracies  of both the  $e_g$ and $t_{2g}$ are 
lifted as in the  ab initio DOSs.  The widths of  $e_g$ and $t_{2g}$ orbitals  are 
also in qualitative agreement with the ab initio results. However, the octahedron 
distortion alone shows that the $t_{2g}$  and $e_g$ do not mix as shown by the color code of the eigenvector 
characters or the five $d$ orbitals, whereas our ab initio DOSs show  small mixing. 
These mixing are due to additional interactions of the iron $d$ states with the other orbitals of the phen atoms. 
Finally, according to the model and our ab initio calculation,
all spin up  $e_g$ and $t_{2g}$ states and  the lowest spin down  $t_{2g}$ state  are occupied, producing an  
iron magnetic moment of  4$\mu_{\rm B}$ in good agreement with our ab initio data shown in table \ref{tab:table}.

\section{Energy barrier and minimal energy path.} 
\begin{figure}[!ht]
	\centering{
\begin{tabular}{c c}
	\includegraphics[scale=0.42] {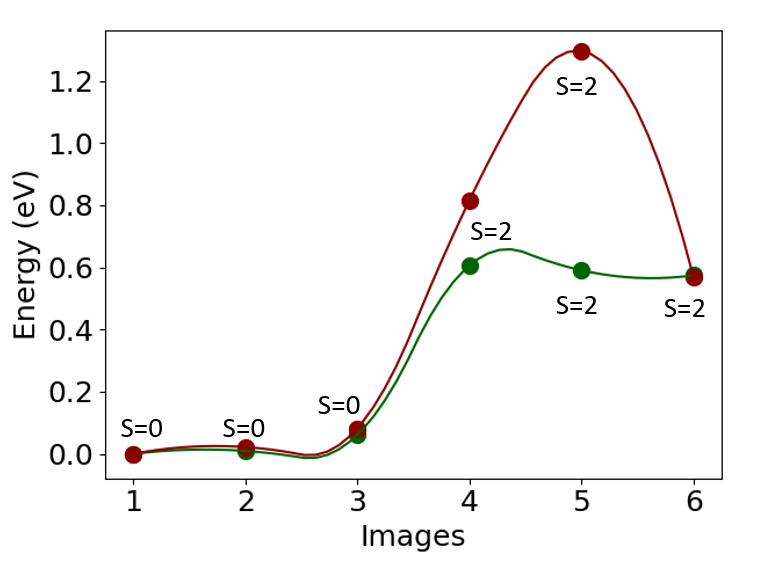}  &
         \includegraphics[scale=0.42]{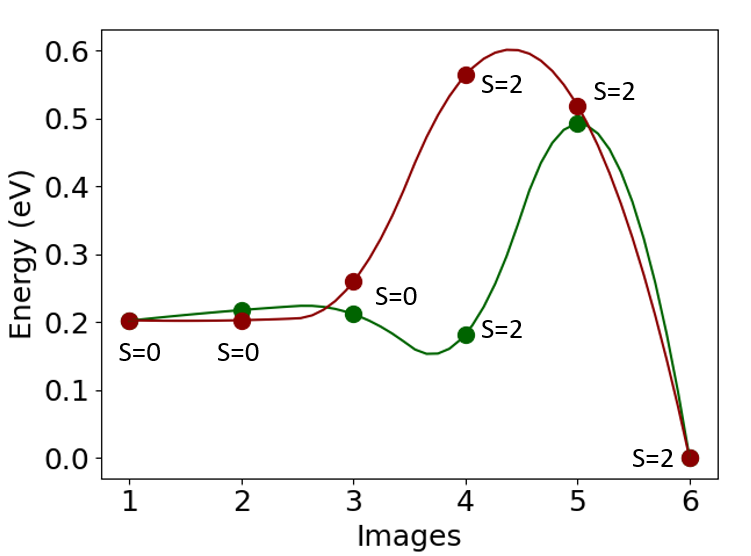} 
\end{tabular}
    }
\caption{ HS-LS minimal energy path and transition barrier for the unfluorinated (left) and the fluorinated (right) 
Fephen molecule adsorbed on Cu(001) surface. The constrained minimization method (see text) are in dark green 
	and the NEB method in dark red. The spin state $S$ of each image is also shown.
The continuous curves are quadratic interpolations used as  guides to the eye.}
\label{fig:neb} 
\end{figure}

To determine the minimal energy path (MEP) between the two magnetic states of
the molecule adsorbed on Cu(001), we start by interpolating linearly four 
images between the coordinates of the HS and LS states.  We then used two
different methods to search for the MEP.  The first one (left in Figure
\ref{fig:neb}), that we call the constrained minimization method (CMM), and the
second method (right of  Figure \ref{fig:neb}) is the aforementioned Nudged
elastic Band (NEB) method \citep{Berne1998}. 
The CMM   minimizes the total energy of each image independently from the others
but constrains  the distance $d_{\rm SS}$ between the two sulfur atoms. This distance 
is interpolated linearly 
from HS state to the LS state and  kept fixed during the atomic relaxation of 
each image.  As a consequence the angle
between the two isothiocyanate (NCS) groups is fixed to the interpolated value.
This constraint doesn't  allow the molecule to return to its HS or LS ground state
and hence gives an estimate of the minimal energy path. 
We don't expect that this method will produce the optimal MEP and hence lowest  energy barrier but
sets a reasonable upper limit.  
In the second method, shown in dark red, we relaxed the entire molecule using the well known
NEB method \cite{Berne1998}.

In principle the NEB method will produce the optimal MEP and energy
barrier, however in practice  it  very difficult to converge  as 
the method is  CPU very intensive for materials with too many atoms per supercell. 
For the unfluorinated molecule adsorbed on
Cu, the NEB method did not converge well as shown in Fig. \ref{fig:neb}. It is surprising
that for the fluorinated molecule the  NEB computed energy barrier
of about 0.56 eV  is slightly higher than  the CMM barrier of 0.49 eV and  produced
the same spin states $S$  along the path as the CMM. This might also indicate that the NEB results are
not fully converged. It is interesting to notice  that for both the fluorinated and unfluorinated
molecule the images near the LS state seem to converge better for the two methods compared to 
images near the HS state.  However, as seen in Fig. \ref{fig:neb} the CMM results are slightly  scattered compared
to those obtained using the NEB method. The reason is that each image in
the CMM is almost independent whereas in the NEB method the images are
connected by spring constants.  It is however  interesting to notice that 
these two independent methods provided roughly the same energy barrier. 

\section{Constant current STM images.} 
Constant current STM studies of Fephen were conducted experimentally, and results were reported for the
H-Fephen molecule on Cu(001) \cite{Gruber2017}.
The authors described the images as formed by  two
types of two oval lobes per molecule. To identify the spin state of each
conformation, scanning tunneling spectroscopy (STS) showed a Kondo peak in
the HS state (where the lobes of the molecule are well separated). In our work, we used the 
Tersoff-Hamann model to construct constant current STM images for the fluorinated
and unfluorinated molecule adsorbed on  Cu(001). To compute the STM images, we  integrated the   local
density of states (LDOS) in a window of $50$  meV around the Fermi level.
In our approximation a  constant intensity will  correspond to a given LDOS. To compare the results of HS and LS
we have chosen arbitrarily the highest LDOS at $4$ {\AA} of the unfluorinated molecule in its LS state 
which is  3.88 $\times 10^{-5}$ electron/\AA$^2$ and 2.44 $\times 10^{-4}$ electron/\AA$^2$ for HS state.  
We have then used these 
values to scan the molecule in its HS and LS 
and determine its corrugation over the surface. 
\begin{figure}[!ht]
	\centering{\includegraphics[scale=0.66]{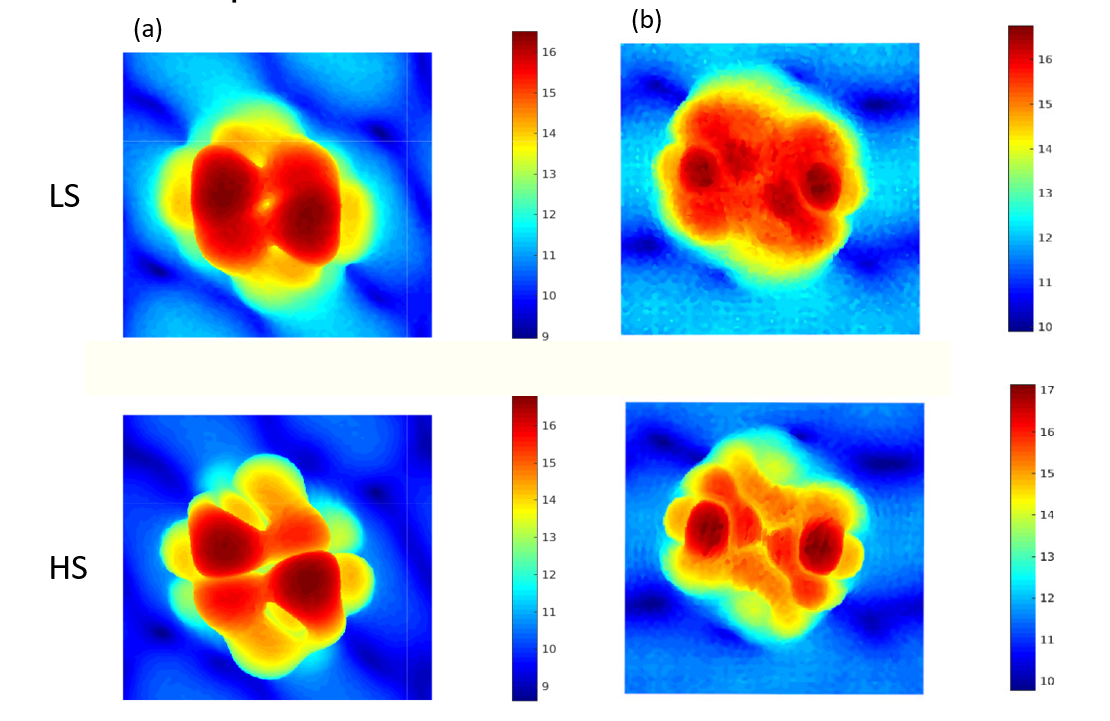}} %
\caption{H-Fephen (a), F-Fephen (b) constant current STM images on copper substrate. 
The top panel gives the  LS state, and the bottom panel the HS state.}
\label{fig:stm}
\end{figure}
Using this procedure, we reproduced the general trend of  the experimental observations for
the H-Fephen molecule. We can see in Figure \ref{fig:stm} (left) that the two oval lobes 
corresponding to the phen groups are  more separated in the HS state and that the hight in the HS state 
is much higher than that of LS state as observed in experiment (see Fig 3b of Ref. \cite{Gruber2017}).
This is because when  we  scan the surface, the current at the same distance from the
molecule ($4$ \AA) is more important in the HS state. For the F-Fephen, the same
conformations were observed although the lobes are more 
expanded in the HS case compared to the unfluorinated molecule. 



\section{Conclusion.}
The substitution of  hydrogen by fluorine in the spin crossover Fephen molecule led to a
profound transformation of its electronic structure. It was shown that the fluorination leads to 
(1) the inversion of the magnetic state for the free and adsorbed
molecule on Cu(001) surface, (2) a noticeable modification of the bond lengths and angles
between the ligands and the central iron atom - though the overall geometry remains unaffected, 
(3) the strong reduction of the $e_g$-$t_{tg}$ ligand-field energy splitting,
especially in the HS state, and (4) a huge reduction of the HS-LS total energy, and a  small reduction of the 
HS to Ls energy barrier.

Finally, the HS and Ls STM images of the fluorinated molecule on Cu(001) surface, calculated using the
Tersoff and Hamann approximation, are very different from those of the unfluorinated molecule. Those results
are of importance since they can be used by future 
experiments to distinguish  the fluorinated molecule from the unfluorinated one and differentiate the HS and LS 
 states.

\subsection{Methods}

To compute the electronic structure and total energy of the SCO molecule adsorbed on a Cu (001) surface, we
used the projected augmented wave (PAW) method \cite{Blochl1994} as
implemented in \textsc{VASP} (Vienna Ab initio Simulation Package) \cite{Kress1996}. The
exchange correlation energy is described within  the generalized gradient
approximation (GGA) \cite{PBE1996}. The long range Van der Waals interaction
between the molecule and the substrate is modeled by means of  the Grimme
approximation \cite{Grimme}. For the description of the localized  $d$ states
of iron, we used the rotationally invariant  DFT+U method as implemented by
Dudarev and coworkers \cite{Dudarev}. 
The values of the Hubbard parameter $U$ and exchange parameter $J$ 
were set respectively to 3 eV and 0.9 eV. These values are shown to reproduce the iron spin
moment and the expected energy difference between the two spin states of Fe(II) \cite{Gueddida2013}.
The electron analysis on each atom  is done with the
Bader electron analysis method \cite{Henkelman2006}. The computation of the
energy barrier and the minimum energy path (MEP) between the high spin and low
spin states is done using 
the nudged elastic band (NEB) method \cite{Berne1998}. 
This method consists in first interpolating many possible
transition points, often called images, from the initial to the final
states and relaxing them with constraints originating from fictitious spring
forces between neighboring images, in order to maintain a constant distance
between them along the reaction path. The calculations converge when the MEP is
found.   But, since we found that
this method is difficult to converge and is computationally prohibitive for 
systems with many atoms per unit cell, we have also used what we called 
 the constrained minimization method (CMM).
In this CMM   we interpolated linearly  the distance  between the two sulfur atoms 
to create different images, we fixed the sulfur-sulfur distance and let 
all the other atoms relax to find the minimal energy path. Surprisingly, as it can be shown
latter, this method led to a slightly smaller energy barrier than the NEB. 

The Fephen molecule contains 51 atoms, where the iron atom is surrounded by 6
nitrogen atoms, forming a distorted octahedron. In addition the molecule is composed of 26 carbon atoms and 16
hydrogen atoms, which will be later substituted by 16 fluorine atoms as indicated in Fig. \ref{fig:relaxed}.  To
simulate the  substrate we have used three layers of copper grown along the
(001) direction, each plane containing 36 (6x6) atoms of copper.  Thus we used
159 atoms in the supercell to fully describe the H(F)-Fephen/Cu(001) system.
Our calculations were done using a periodic cell of ($17.8\times17.8\times33.6$
\AA) for both the  gas phase and the molecule adsorbed on the copper
substrate.

The total energy is converged to  $ 10^{-5}$ eV and the  plane waves cut-off energy is
set at $ 500$ eV. The atoms are allowed to move until the forces are below $ 10^{-3}$
eV/\AA\; in each direction of the cell axes.  
The calculation of the Fermi level is carried with a Gaussian integration
method \cite{Methfessel1989}, using a width of  $0.1$ eV. This electronic
entropy is removed when calculating  the total energy.  Due to the large number
of atoms per supercell, we restricted our ${\bf k}$-point mesh to the $\Gamma$
point. 
\begin{acknowledgement}
This work of the Interdisciplinary Thematic Institute QMat, as part of the ITI 2021 2028 program of the University of Strasbourg, CNRS and Inserm, was supported by IdEx Unistra (ANR 10 IDEX 0002), and by SFRI 
STRATUS project (ANR 20 SFRI 0012) and EUR QMAT ANR-17-EURE-0024 under the framework of the French Investments for the Future Program.
This work was performed using HPC resources from GENCI-CINES Grant gem1100 and the HPC supercomputer of the
University of Strasbourg.
\end{acknowledgement}



~\\
\noindent \textbf{Note:} The authors declare no competing financial interest.


\bibliography{kha}
\end{document}